# Myotubularin MTM1 Involved in Centronuclear Myopathy and its Roles in Human and Yeast Cells

Dimitri L. Bertazzi#, Johan-Owen De Craene# and Sylvie Friant*

*Department of Molecular and Cellular Genetics, UMR7156, Université de Strasbourg and CNRS, France*

#Authors contributed equally to this work.

*Corresponding author: Friant S, Department of Molecular and Cellular Genetics, UMR7156, Université de Strasbourg and CNRS, 67084 Strasbourg, France, E-mail: s.friant@unistra.fr

## Abstract

Mutations in the MTM1 gene, encoding the phosphoinositide phosphatase myotubularin, are responsible for the X-linked centronuclear myopathy (XLCNM) or X-linked myotubular myopathy (XLMTM). The MTM1 gene was first identified in 1996 and its function as a PtdIns3P and PtdIns(,5)P2 phosphatase was discovered in 2000. In recent years, very important progress has been made to set up good models to study MTM1 and the XLCNM disease such as knockout or knockin mice, the Labrador Retriever dog, the zebrafish and the yeast *Saccharomyces cerevisiae*. These helped to better understand the cellular function of MTM1 and of its four conserved domains: PH-GRAM (Pleckstrin Homology-Glucosyltransferase, Rab-like GTPase Activator and Myotubularin), RID (Rac1-Induced recruitment Domain), PTP/DSP (Protein Tyrosine Phosphatase/Dual-Specificity Phosphatase) and SID (SET-protein Interaction Domain). This review presents the cellular function of human myotubularin MTM1 and its yeast homolog yeast protein *Ymr1*, and the role of MTM1 in the centronuclear myopathy (CNM) disease.

## Introduction

The X-linked centronuclear myopathy (XLCNM) or X-linked myotubular myopathy (XLMTM) was first described in 1966 by Spiro and coll. and named myotubular myopathy [1]. A year later, this disease was also described as a centronuclear myopathy [2]. This myopathy, affecting 1 in 50000 male births, is caused by mutations in the MTM1 gene (OMIM® ID 310400) coding for a phosphoinositides phosphatase [3,4]. The responsible gene was localized on the X-chromosome Xq28 region in 1990 [5] and first identified by Laporte and coll. in 1996 [6]. The MTM1 gene is 104 749 bp in length and consists of 15 exons [7]. MTM1 is ubiquitously expressed since its 3.9 kb transcript was detected in many tissues including hart, brain, placenta, liver, muscles and pancreas [6,7]. However, the disease specifically affects skeletal muscles but no muscle specific exon or splicing variant has been detected to explain this selectivity [6].

MTM1 mutations are subjected to an X-linked recessive mode of inheritance. Indeed, the mother, usually asymptomatic, is the heterozygous carrier of the MTM1 mutations [8]. To date, more than 200 different mutations have been identified in the MTM1 gene [8-14]. The majority of mutations are nonsense or frameshift mutations leading to the absence of MTM1 or to the production of a truncated non-functional protein. Missense mutations have also been described and are responsible for mild to severe forms of myopathy (Figure 1). Interestingly, these different mutations are distributed throughout the MTM1 gene and affect the various domains of the MTM1 protein but lead to the same disease with various degrees of severity (Figure 1).

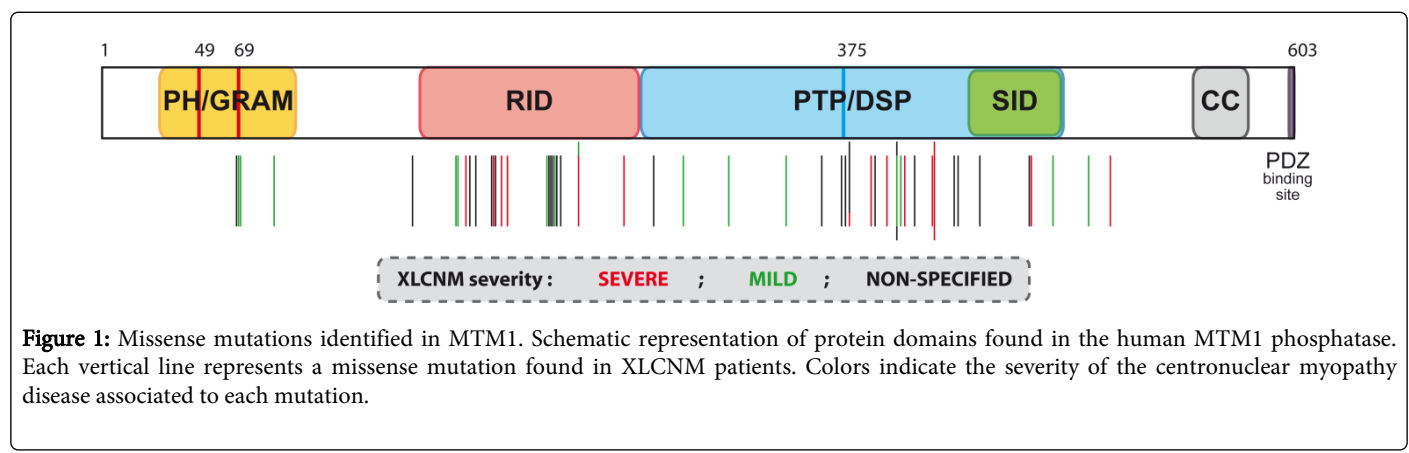

**Figure 1:** Missense mutations identified in MTM1. Schematic representation of protein domains found in the human MTM1 phosphatase. Each vertical line represents a missense mutation found in XLCNM patients. Colors indicate the severity of the centronuclear myopathy disease associated to each mutation.

## Clinical and Histological Features of MTM1 Myopathy

XLCNM was first identified in 1966 based on the clinical study of a 12-year-old boy presenting a general muscular weakness, ophtalmoplegia and ptosis [1]. Muscular biopsies showed characteristics of immature muscles with hypotrophic fibers and centrally located nuclei that resembled fetal myotubes, leading the disease to be named myotubular myopathy. The disease was believed to result from a defect in muscle maturation. This was later supported by studies showing that Vimentin and the prenatal myosin heavy chain (MHC) isoform, two proteins specific of fetal myotubes, were detected in muscles of XLCNM patients [15,16]. However, the study of Mtm1 knockout (KO) mice revealed that these phenotypes result from an inability to maintain muscles in their mature state. The Mtm1 KO mice show clinical signs of the myopathy starting around 4 weeks of age and most die in the first 2 months from progressive and generalized muscle disorders having the signs of XLCNM [17]. The etiology of the disease has been validated by other studies, despite the fact that markers of immature muscle are not always present in XLCNM patients [18].

XLCNM is characterized by generalized muscle weakness and respiratory distress in affected males whereas female heterozygous carriers are generally asymptomatic. The disease ranges from severe to mild with some severe forms leading to neonatal death due to respiratory failure. In healthy mature muscles, nuclei are localized at the periphery of the fiber close to the plasma membrane, whereas in XLCNM muscles, many nuclei are centrally localized. The centrally located nuclei are generally surrounded by a halo lacking contractile elements but containing mitochondria aggregates and glycogen granules [19].

## The Myotubularin Family of Protein

Myotubularins belong to the PTP/DSP (Protein Tyrosine Phosphatase/Dual-Specificity Phosphatase) super-family. MTM, the founding member of the myotubularin family, dephosphorylates the D3 phosphate of the inositol ring of two types of phosphoinositides (PIP): the phosphatidylinositol 3-phosphate (PtdIns3P) and the phosphatidylinositol 3,5-bisphophate (PtdIns(3,5) P2) [3,4,20,21].

Myotubularin homologues are found in all eukaryotic cells, from Plant to Fungi and Human (Table 1) [22,23]. Most eukaryotes have several myotubularin isoforms, however in the yeast *Saccharomyces cerevisiae* there is only one named *Ymr1* (Yeast myotubularin related 1) despite its whole-genome duplication [2,24]. In Human, there are 14 paralogues of myotubularin and one closely related phosphatase termed Jumpy. The high degree of conservation among members of the myotubularin family suggests they arose by gene duplication and perform important cellular functions in different cell types [2,23]. Myotubularins are PIP phosphatases, but among the 14 human myotubularins some members do not possess the cysteine or the arginine residue of the CX5R PTP/DSP conserved catalytic motif rendering these phosphatases catalytically inactive. This allows the classification of myotubularins in either active phosphatases: MTM, MTMR1, MTMR2, MTMR3 (FYVE-DSP1), MTMR4 (FYVE-DSP2), MTMR6, MTMR7 and MTMR8 and, inactive phosphatases: MTMR5 (Sbf1), MTMR13 (Sbf2), MTMR9 (LIP-STYX), MTMR10 (FLJ20313), MTMR11 (CRA) and MTMR12 (3-PAP) [25]. Besides MTM1 mutations affecting muscle, defects in the active MTMR2 or the catalytically inactive MTMR13 myotubularins result in a neuropathy, the Charcot-Marie-Tooth syndrome [26,27].

| Organism | Number of paralogues | Paralogues |
|---|---|---|
| *Homo sapiens* | 14 + 1 | MTM1, MTMR1-13, hJumpy |
| *Mus musculus* | 14 + 1 | MTM1, MTMR1-13, mJumpy |
| *Canis lupus* | 13 | MTM1, MTMR1-7, MTMR9-13 |
| *Danio rerio* | 14+1 | MTM1, MTMR1-13, dJumpy |
| *Drosophila melanogaster* | 7 | Mtm, MtmR2/3, MtmR3/4, MtmR6/7/8, MtmR9, MtmR10/11/12, MtmR3/13 |
| *Caenorhabditis elegans* | 5 | MTM-1, MTM-3, MTM-5, MTM-6, MTM-9 |
| *Saccharomyces cerevisiae* | 1 | YMR1 |
| *Arabidopsis thaliana* | 2 | AtMTM1, AtMTM2 |

**Table 1:** Number of myotubularins found in various organisms.

## The Yeast Ymr1 Myotubularin

In yeast *S. cerevisiae*, deletion of the *YMR1* gene coding for the only myotubularin homologue is not lethal for the cell. Nonetheless, *ymr1Δ* knockout cells have fragmented vacuoles (the yeast lysosome) (Figure 2) and a two-fold increase in intracellular levels of PtdIns3P [4,28,29].

The Ymr1 protein shares the regulation of intracellular PtdIns3P levels with synaptojanin-like phosphatases Sjl2/Inp52 and Sjl3/Inp53. Indeed, the triple deletion of *YMR*, *SJL2* and *SJL3* is lethal. However, Sjl2 and Sjl3 phosphatases are less specific than Ymr1 since they also dephosphorylate D4 and D5 positions of the inositol ring, whereas Ymr1 dephosphorylates only at position D3 [30]. In wild-type yeast cells, *YMR1* overexpression induces a protein trafficking defect from the Golgi apparatus towards the vacuole along the VPS (vacuolar protein sorting) pathway [28]. Moreover, the *ymr1Δ sjl3Δ* double mutant strain has defects in intracellular trafficking at endosomes [28]. The Ymr1 myotubularin is also required for autophagy since the phosphatase-dead Ymr1-C397S catalytic mutant is impaired in autophagy and the Sjl3 synaptojanin cannot substitute Ymr1 for this function [31]. Electron microscopy analyses under autophagic conditions show that autophagosomes bearing several Atg proteins accumulate in the cytoplasm of *ymr1Δ* mutant cells. In wild-type cells under these autophagic conditions, Ymr1 localizes to the phagophore assembly site (PAS) suggesting its phosphatase activity could be required to regulate the PtdIns3P turnover necessary for autophagosome maturation [31].

The fragmented vacuolar phenotype and the PtdIns3P turnover defects observed in yeast *ymr1Δ* mutant cells are restored by expressing the human MTM1 myotubularin (Figure 2), showing that yeast is a model well suited to study MTM1 [29]. Moreover, the catalytic activity of MTM1 is required to complement the vacuolar defect of the *ymr1Δ* mutant cells (Figure 2) [29]. Therefore, yeast can be used to determine the in vivo phosphatase activity of different MTM1 patient mutants with an easy assay based on the vacuolar size.

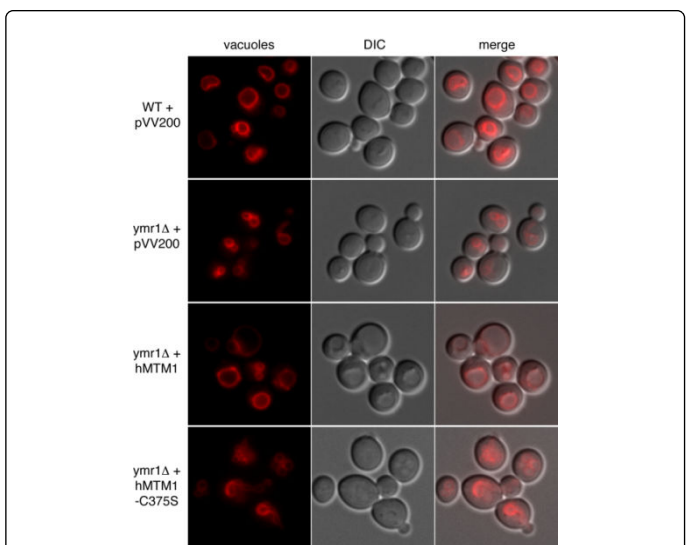

**Figure 2:** Human MTM1 complements the fragmented vacuolar phenotype of the yeast *ymr1Δ* mutant cells. Yeast wild-type (WT) or *ymr1Δ* mutant cells transformed with an expression plasmid (pVV200) bearing wild-type human MTM1 or the catalytic-dead MTM1-C375S mutant were grown to exponential phase and the vacuoles were stained by FM4-64. Cells were observed by fluorescence microscopy with DIC (Nomarski) and TRITC (FM4-64) filters. Images were treated with ImageJ software (Rasband, W.S., ImageJ, U. S. National Institutes of Health, Bethesda, Maryland, USA, 1997–2009).

## The Human Myotubularin MTM1

The human MTM1 protein is 603 amino acids in length with 4 domains: PH-GRAM (Pleckstrin Homology-Glucosyltransferase, Rab-like GTPase Activator and Myotubularin), RID (Rac1-Induced recruitment Domain), PTP/DSP and SID (SET-protein Interaction Domain) (Figure 3). All human paralogues and the yeast Ymr1 myotubularins share these 4 domains. Besides these core domains, MTM1 has also a C-terminal coiled-coil motif followed by a PDZ binding domain but no transmembrane domain.

## The PH-GRAM Domain

The PH-GRAM domain is at the N-terminus of the MTM1 protein (amino acids 29 to 160). Its name derives from its homology to the PH (Pleckstrin Homology) and GRAM (Glucosyltransferase, Rab-like GTPase Activator and Myotubularin) domains (Figure 1). The PH-GRAM domain is present in almost all eukaryotic myotubularins, suggesting that its architecture was established early in evolution, perhaps coinciding with the acquisition of the characteristic PTP/DSP myotubularin phosphatase domain [23].

The PH domain was initially found in a platelet and leucocyte protein named Pleckstrin (Platelet and Leucocyte C Kinase Substrate). The Pleckstrin PH domain comprises between 100 and 120 amino acids organized in an up-and-down β-barrel of seven antiparallel β-strands followed by a C-terminal α-helix [32]. It binds membrane lipids such as phosphoinositides but also recruits effectors and/or regulatory proteins. Hundreds of proteins with a PH domain have been identified in human cells and display a great variability, especially in the loops connecting the β-strands. All PH domains bind phosphoinositides, but they do not have the same specificity and affinity for their substrate(s). Indeed, it is estimated that only 10% of the PH domains bind PIPs with high affinity and specificity [33].

The GRAM domain was identified by Doerks and coll. in a number of human proteins such as myotubularins [34]. Since this GRAM domain displays high structural similarities with the PH domain, it was renamed PH-GRAM domain. Like PH domains, PH-GRAM domains also bind to both PIPs and effector proteins. The affinity of a PH-GRAM domain for a given PIP could play a key role in targeting the various myotubularins to a specific intracellular compartment. Indeed, PH-GRAM domain amino acids sequence can be very divergent, suggesting that these domains could have distinct molecular properties and cellular functions [23]. The PH-GRAM domain of MTM1 has the highest affinity for the endosomal enriched phosphoinositide PtdIns(3,5)P2 [35]. Different XLCNM patient mutations affecting the PH-GRAM domain of MTM1 were tested for PIP interaction, the V49F shows a significant decrease in PtdIns(3,5)P2 binding, whereas the R69C, L70P or L87P have only a minor effect on PtdIns(3,5)P2 binding [35]. Interestingly, the MTM1-V49F mutant, responsible for a severe form of myopathy, shows a twofold decrease in phosphatase activity compared to the MTM1-R69C mutant, associated with milder forms of myopathy, which displays wild-type phosphatase activity [29]. The crystal structure of MTMR2 (PDB accession number 1ZSQ) was used to model the MTM1 protein and the model shows that residue R69 is at the MTM1 protein surface, whereas residue V49 is more buried in the interior of the protein (Figure 3). This suggests that mutation of this V49 residue might be more deleterious for the overall structure of the MTM1 PH-GRAM domain.

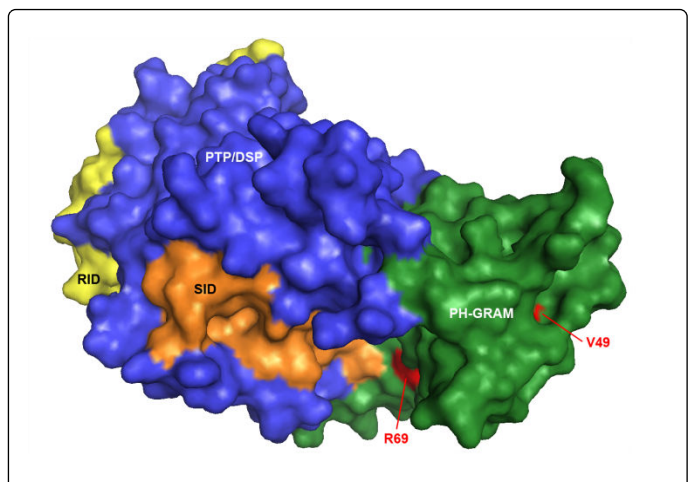

**Figure 3:** Model of the MTM1 protein with the different domains. This MTM1 protein model is based on the MTMR2 crystal structure (PDB accession number 1ZSQ) [29]. The different domains are shown in different colors: PH-GRAM (orange), RID (yellow), PTP/DSP (blue), SID (green). The V49 and R69 residues are highlighted in red. The figure was prepared with the PyMOL software (The PyMOL Molecular Graphics System, Version 1.5.0.1 Schrödinger, LLC.).

## The RID Domain

The RID domain (amino acids 161-272 of MTM1) was identified for its ability to localize MTM1 at the plasma membrane in cells expressing a constitutively activated Rac1 GTPase mutant [36]. Rac1 belongs to the Rho and Cdc42 GTPase subfamily involved in plasma membrane ruffling and endocytosis through actin cytoskeleton remodeling [37]. Expression of the dominant active mutant Rac1-V12 induces widespread plasma membrane remodeling and MTM1 localization to these ruffles [36]. Mutations or small deletion into the RID domain prevent MTM1 localization to these Rac1-V12 induced membrane ruffles, showing that the RID domain is crucial for its plasma membrane localization [36]. Moreover, the RID domain of MTM1 also interacts with Desmin, an intermediary filament protein involved in shaping muscle cells [38].

The RID domain plays a crucial role in the recruitment of MTM1 to plasma membrane ruffles after Rac1 GTPase activation but this membrane localization could also be enhanced in the muscle cells through its interaction with the actin cytoskeleton via Desmin.

## The PTP/DSP Phosphatase Domain

The lipid phosphatase PTP/DSP (amino acids 273-471 of MTM1) domain is also highly conserved among the different myotubularins. The catalytic site of the PTP/DSP domain consists of a highly conserved CX5R motif (VHCSDGWDRT) that defines the active myotubularin subfamily [25]. It was shown that three residues of this catalytic site are important: the cysteine and arginine residues to stabilize the substrates by forming a thiol-phosphate intermediary and the aspartate involved in the product release.

Initially described as a protein tyrosine phosphatase (PTP) due to the presence of the CX5R motif characterizing this type of enzyme, it was later shown that the PTP/DSP domain of MTM1 dephosphorylates PtdIns3P and PtdIns(3,5)P2 in PtdIns and PtdIns5P respectively [3,4,21]. This particular PIP substrate preference is no exception, since the human phosphatase PTEN (Phosphatase and TENsin homologue), also characterized by a CX5R motif, displays a substrate preference towards PtdIns (3,4,5) P3 [39].

In keeping with the high degree of conservation of the myotubularin PTP/PSD domain, myotubularins carrying the CX5R motif are enzymatically active and dephosphorylate PtdIns3P and PtdIns(3,5)P2 in vitro [21]. In vivo, in yeast cells or in the mouse muscle, residue C375 of the CX5R motif is also necessary for MTM1 phosphatase activity [29].

## The SID Domain

The SID domain (amino acids 435-486 of MTM1) was identified in 1998 by Cui and coll. in the inactive myotubularin Sbf1 (SET-binding factor)/MTMR5 [40]. Contrary to the other domains found in myotubularins, this domain is specific to the myotubularin family. It was shown that the SID domain binds to proteins containing the SET domain (Suvar3-9, Enhancer of zeste, Tritorax) found in proteins involved in epigenetic regulation of genes [41].

Recombinant MTM1 and MTMR5/Sbf1 were shown to bind the yeast methyltransferase Set1 SET domain in vitro and the human histone methyltransferase SET domain in vivo [40]. Overexpression of the MTMR5/Sbf1 inhibits the C2 myoblast fusion to form myotubes, and the SID domain is required for this inhibition [40]. Thus the interaction between the SID domain of myotubularins and proteins with a SET domain could play an important role in regulating the expression of some genes.

## Study of XLCNM in Different Animal Model

Different models are used to better understand the XLCNM pathology, among them the zebrafish, mouse and Labrador Retriever are the ones mimicking best the different hallmarks of this centronuclear myopathy. The zebrafish *Danio rerio* has one MTM1 homologue and 12 of the 14 human MTMR genes homologues (Table 1). The antisens morpholino strategy was used to knockdown the MTM1 gene and the resulting MTM morphants had severe muscle pathologies with mislocalized nuclei, myofibers hypotrophy and T-tubule disorganization [42].

The best studied XLCNM model is the Mtm1 knockout mouse model generated by Buj-Bello and coll. [17]. This Mtm1 KO mouse model is viable but its lifespan is severely reduced and it shows all the XLCNM phenotypes with muscle weakness, muscle fibers hypotrophy and accumulation of central nuclei [17]. The XLCNM like phenotype of these Mtm1 KO mice can be corrected by adeno-associated virus AAV-mediated intramuscular injection of wild-type MTM1 [43]. Interestingly, the centronuclear myopathy phenotypes of these Mtm1 KO mice are also rescued by a catalytic-dead MTM1-C375S mutant or by an inactive MTM1-S376N mutant found in patients, suggesting that the lack of catalytic phosphatase activity is not the main cause of the disease [29]. However, these different gene therapies were based on muscular injection of AAV-MTM1 and could only determine whether the histological features of the XLCNM disease were corrected. Recently a new therapy based on the systemic delivery of the MTM1 cDNA under the control of the muscle-specific desmin promoter was tested in the Mtm1 KO mice [44]. The distinctive histological features of the XLCNM disease (centronuclear nuclei, aberrant accumulation of mitochondria and T-tubules misorganization) are restored and the gene therapy also prolongs the survival and the muscular strength of the treated Mtm1 KO mice [44]. These promising results pave the way to a clinical trial of gene therapy for XLCNM. Another model mouse was developed called Mtm1 pR69C, which is a knockin of the MTM1-R69C patient mutation. These Mtm1 pR69C mice display reduced levels of MTM1-R69C transcripts and facilitate the study of XLCNM since they present less severe symptoms of the disease [45]. Recently, Mtm1 KO and Mtm1 pR69C mice were used to test a new therapy based on intramuscular injection of a muscle-specific monoclonal antibody fused N-terminally to the MTM1 protein [46]. This strategy leads also to an improvement of the XLCNM-associated phenotypes, showing that either gene therapy or exogenous protein supplementation therapy show promise for treatment of the XLCNM disease.

The last model for XLCNM is the Labrador Retriever, a very important model for testing new therapeutic strategies because it is a large animal model that best mimics the human pathology and that the centronuclear myopathy cnm mutation results from a spontaneous genetic abnormality (mutation MTM1-N155K) [47,48]. Recently a systemic MTM1 AAV-mediated therapy showed that not only histological features of the XLCNM disease could be rescued by AAV-MTM1 injection in the Labrador Retriever cnm mutant dogs, but that this gene therapy also led to prolong survival associated with improved muscular strength, locomotion and respiratory functions [44].

## Conclusion

To date, more than 200 different mutations linked to centronuclear myopathy have been identified in the MTM1 gene and these mutations are present in the different domains of the MTM1 protein (Figure 1). All these mutations lead to the same disease with only the severity varying. Moreover, many mutations responsible for XLCNM affect amino acid residues that are conserved throughout evolution, suggesting that these residues could be important for the cellular function of MTM1.

The identification of mutations in the MTM1 gene as the sole cause of the disease has allowed for the pursuit of gene therapy approaches to treat the associated centronuclear myopathy disease. To this effect, a systemic injection of AAV-MTM1 in Mtm1 KO mice was successful and conferred long-term survival of mice that displayed muscle robustness and no centronuclear myopathy histological features [44]. Since the XLCNM etiology in mice is somewhat different than in humans, the authors also used the Labrador Retriever to test this promising therapy and obtained a similar success as in mice [44].

Whereas therapeutic research has made great progress, progress in fundamental research on the disease and on myotubularins in general is much slower. Indeed, the link between this phosphatase activity and the disease is only poorly understood since mutations are present not only in the catalytic PTP/DSP domain but also in the other MTM1 domains (Figure 1). Moreover some patient mutations responsible for severe forms of the disease display wild-type catalytic phosphatase activity [29]. These results were acquired through the use of yeast *S. cerevisiae* as a model because it allows the in vivo analysis of a myotubularin in the absence of its numerous paralogues (Table 1). These findings suggested that XLCNM Mtm1 KO mice phenotypes could be ameliorated by AAV-mediated muscular expression of the MTM1-C375S phosphatase-dead myotubularin, which was indeed observed [29]. This unexpected finding raises the question about the involvement of the PIP phosphatase activity of MTM1 in the disease and the other functions MTM1 carries out. However, this study has to be confirmed by a systemic expression of AAV-MTM1-C375S to show that the lifespan and muscular strength of Mtm1 KO mice are ameliorated. Since it seems that the phosphatase activity is only marginally involved in the disease, this suggests that MTM1 is involved in a different interaction network than are the other myotubularins. In this regard we know that mutations in the amphiphysin BIN1 and the dynamin DNM2 also cause centronuclear myopathies but the link with MTM1 is still unclear [49]. The results found in recent papers strongly underline the power interplay between *S. cerevisiae*, the mouse and the dog models would have on the better understanding of why mutations in this ubiquitously expressed MTM1 gene are responsible for a muscle specific disease. This is also true for other questions of interest such as why MTM1 mutations are not compensated by other closely related active myotubularins such as MTMR1 or MTMR2 also present in the muscles [25].

It is abundantly clear that studying the more tractable XLCNM will benefit the study of other myotubularin diseases such as the Charcot-Marie-Tooth syndrome involving either MTMR2 or MTMR13. And more broadly, understanding these two very different myotubularins will help us understand the complex functions and regulations of the other myotubularins.

## Acknowledgments

The authors would like to thank Jean-Louis Mandel and Jocelyn Laporte (IGBMC, Illkirch, France) for their constant support throughout the work and Bruno Klaholz (IGBMC, Illkirch) for the structural model of the MTM1 protein. This work was supported by the CNRS (ATIP-CNRS 05-00932 and ATIP-Plus 2008-3098 to S.F.), the Fondation Recherche Médicale (FRM INE20051105238 and FRM-Comité Alsace 2006CX67-1 to S.F. and FRM Postdoctoral fellowship to J-O.D.C.), the Association Française contre les Myopathies (AFM-SB/CP/2013-0133/16551 grant to SF and fellowship to DLB) and the Agence Nationale de la Recherche (ANR-07-BLAN-0065 and ANR-13-BSV2-0004).